\documentclass[smallextended]{svjour3}
\usepackage{natbib}
\usepackage{graphicx}
\usepackage{amsmath}
\usepackage{multirow}
\usepackage{booktabs}
\usepackage{mathptmx}

\journalname{Experimental Astronomy}
\title{Noise statistics in a fast digital radio receiver: the Bedlam backend for the Parkes Radio Telescope}
\titlerunning{Noise statistics in the Parkes Bedlam backend}

\author{J.D. Bray \and R.D. Ekers \and P. Roberts}

\institute{
 J.D. Bray
  \at School of Chemistry \& Physics, University of Adelaide, Australia\\\email{justin.bray@adelaide.edu.au}
\and
 J.D. Bray \and R.D. Ekers \and P. Roberts
  \at CSIRO Astronomy \& Space Science, Epping, Australia
}

\date{Received: date / Accepted: date}

\begin{document}
\def\arxiv{ArXiv e-prints }    
\def\aspacer{Adv.~Space R. }   
\def\aap{A\&A }                
\def\aapr{A\&A~Rev. }          
\def\aaps{A\&AS }              
\def\aj{AJ }                   
\def\ajph{Australian J.~Phys. }
\def\alet{Astro.~Lett. }       
\def\anchem{Analytical~Chem. } 
\def\ao{Applied Optics }       
\def\apj{ApJ }                 
\def\apjl{ApJ }                
\def\apjs{ApJS }               
\def\app{Astropart.~Phys. }    
\def\apss{Ap\&SS }             
\def\araa{ARA\&A }             
\def\arep{Astron.~Rep }        
\def\aspconf{Astron.~Soc.~Pac.~Conf. } 
\def\asr{Av.~Space Res. }      
\def\azh{AZh }                 
\def\baas{BAAS }               
\def\bell{Bell~Systems~Tech.~J. } 
\def\cpc{Comput.~Phys.~Commun. } 
\def\epsl{Earth and Plan.~Sci.~Lett. } 
\def\expa{Exp.~Astron. }       
\def\gca{Geochim.~Cosmochim.~Acta } 
\def\grl{Geophys.~Res.~Lett. } 
\def\iaucirc{IAU Circ. }       
\def\ibvs{IBVS }               
\def\icarus{Icarus }           
\def\ieeetit{IEEE Trans.~Inf.~Theor. } 
\def\ijmpd{Internat.~J.~Mod.~Phys.~D } 
\def\invp{Inverse Prob. }      
\def\jastp{J.~Atmos.~Solar-Terr.~Phys. } 
\def\jcap{J.~Cosm.~Astropart.~Phys. } 
\def\jcomph{J.~Comput.~Phys. } 
\def\jcp{J.~Chem.~Phys. }      
\def\jewa{J.~Electromagn.~Wav.~Appl } 
\def\jgeod{J.~Geodesy }        
\def\jgr{J.~Geophys.~R. }      
\def\jhep{JHEP }               
\def\jrasc{JRASC }             
\def\met{Meteoritics }         
\def\mmras{MmRAS }             
\def\mnras{MNRAS }             
\def\moonp{Moon and Plan. }    
\def\mpla{Mod.~Phys.~Lett.~A } 
\def\mps{Meteoritics and Planetary Science } 
\def\nast{New Astron. }        
\def\nat{Nature }              
\def\nima{Nucl.~Instrum.~Meth.~A } 
\def\njp{New J.~Phys. }        
\def\nspu{Phys.~Uspekhi }      
\def\pasa{PASA }               
\def\pasj{PASJ }               
\def\pasp{PASP }               
\def\phr{Phys.~Rev. }          
\def\pla{Phys.~Lett.~A }       
\def\plb{Phys.~Lett.~B }       
\def\pop{Phys.~Plasmas }       
\def\pra{Phys.~Rev.~A }       
\def\prb{Phys.~Rev.~B }        
\def\prc{Phys.~Rev.~C }        
\def\prd{Phys.~Rev.~D }        
\def\prl{Phys.~Rev.~Lett. }    
\def\pst{Phys.~Scr.~T }        
\def\phrep{Phys.~Rep. }        
\def\phss{Phys.~Stat.~Sol. }        %
\def\procspie{Proc.~SPIE }     
\def\planss{Planet.~Space Sci. }  
\def\qjras{QJRAS }             
\def\radsci{Radio Sci. }       
\def\rpph{Rep.~Prog.~Phys. }   
\def\rqe{Rad.~\&~Quan.~Elec. } 
\def\rgsp{Rev.~Geophys.~Space Phys. } 
\def\rsla{Philos.~Trans.~R.~Soc.~Lond.~A } 
\def\sal{Sov.~Astron.~Lett. }
\def\spjetp{Sov.~Phys.~JETP }  
\def\spjetpl{Sov.~Phys.~JETP~Lett. } 
\def\spu{Sov.~Phys.~Uspkehi }  
\def\sci{Science }             
\def\solph{Sol.~Phys. }        
\def\ssr{Space Sci.~Rev. }     
\def\wars{Workshop on App.~of Radio Sci. } 
\def\zap{Z.~Astrophys. }       

\newcommand{\expfrac}[3][]{ e^{ #1 #2 / #3 } }

\newcommand{\sinc}{{\rm sinc}}
\newcommand{\sign}{{\rm sign}}
\newcommand{\erf}{{\rm erf}}

\newcommand{\subsc}[1]{\ensuremath{_\textsc{#1}}}
\newcommand{\subLO}{\subsc{lo}}
\newcommand{\subIF}{\subsc{if}}
\newcommand{\subRF}{\subsc{rf}}

\newcommand{\expec}[1]{{\rm E}[#1]}
\newcommand{\var}[1]{{\rm Var}[#1]}

\newcommand{\fgcorr}{\alpha}
\newcommand{\fprimestd}{\varsigma}
\newcommand{\hprimestd}{\varsigma_h}
\newcommand{\fthr}{f_{\rm thr}}
\newcommand{\hthr}{h_{\rm thr}}

\newcommand{\bignum}[2]{#1,#2}

\maketitle

\begin{abstract}
 The digital record of the voltage in a radio telescope receiver, after frequency conversion and sampling at a finite rate, is not a perfect representation of the original analog signal.  To detect and characterise a transient event with a duration comparable to the inverse bandwidth it is necessary to compensate for these effects, altering the statistical properties of the signal and making it difficult to determine the significance of a potential detection.  We present an analysis of these modified statistics and demonstrate them with experimental results from Bedlam, a new digital backend for the Parkes radio telescope.
\end{abstract}

\section{Introduction}
\label{sec:intro}

Typical radio astronomy applications are not sensitive to transitory extreme fluctuations in the signal voltage from the telescope receiver.  For example, when measuring the spectral intensity of a static astronomical object, mere single-bit sampling of the voltage achieves a minimum of 64\% of the signal-to-noise ratio for the ideal many-bit case, even though the ability to measure the magnitude of a single sample is entirely lost~\citep[sec. III]{weinreb1963}.  However, in applications in which a single high-magnitude sample is observationally meaningful --- implying a signal with time structure on the scale of the inverse bandwidth --- it becomes necessary to examine the single-sample behaviour of the telescope receiver, and to develop an understanding of its statistics.

One such application is the search for nanosecond-scale broad-band radio pulses from neutrino-induced particle cascades in the Moon, recently conducted with the WSRT~\citep{buitink2010}, EVLA~\citep{jaeger2010} and ATCA~\citep{james2010} telescopes.  These pulses are inherently very narrow in time, with a detectable width approximately equal to the inverse bandwidth resulting from band-limiting in the radio receiver.  A single such pulse would appear in the voltage recorded by the receiver as a transitory excursion --- possibly only a single sample --- beyond the statistical bounds of the background Gaussian noise, here dominated by thermal radiation from the Moon.  If such an event were detected, and a possible origin as radio frequency interference (RFI) were excluded, it might indicate the registration of an ultra-high energy neutrino, with significant implications for cosmic ray astrophysics~\citep{beresinsky1969}.  The confidence of the detection would depend on the expected rate of false positives from noise in the receiver, which we address in this paper.

Radio receivers typically convert the incoming signal from its original radio frequency to a lower intermediate frequency by mixing it with the single-frequency output from a local oscillator.  This process affects time-averaged properties of the signal, such as the spectral intensity, in a predictable fashion, but the effect on the shape and height of an individual peak will depend on its phase relative to the local oscillator.  Since this phase is usually unknown, the original signal cannot be unambiguously reconstructed; to search for peaks in the original signal, it is necessary to reconstruct all of its possible states, searching in phase for the maximum peak height.  In section~\ref{sec:phase} we illustrate this effect, and show that the signal envelope defined by \citet{longuet-higgins1957} reconstructs the maximum possible peak height.

The fraction of digital samples above some detection threshold can be trivially calculated with Gaussian statistics.  However, a peak in the original analog signal will generally lie between two sampled points, neither of which will record its full magnitude, as first pointed out in this context by \citet{james2010}.  The values of the analog signal between the sampled points, including any such peaks, can be reconstructed through convolution with a $\sinc$ function as a consequence of the Nyquist-Shannon sampling theorem.  The number of expected excursions beyond some threshold in this interpolated signal was derived by \citet{rice1944,rice1945}; in section~\ref{sec:exrate} we explain the application of this result and extend it to find the excursion rate for the signal envelope.

We test the above results through simulation and experiment.  In section~\ref{sec:mc} we perform Monte Carlo simulations of the excursion rates for noise with various spectra, finding them to be consistent with the analytic results.  In section~\ref{sec:bedlam} we present the results of an experiment in which a noise source was connected to ``Bedlam'' --- a new digital backend for the Parkes radio telescope --- which recorded extreme fluctuations in the signal voltage for a period of 20 days.  Bedlam was designed and built for LUNASKA (Lunar Ultra-high energy Neutrino Astrophysics with the Square Kilometre Array), a neutrino detection project of the type described above~\citep{bray2010}.  The experiment described here allows us to test its consistency with analytic and simulated models of the noise behaviour, and to calibrate any variation.

\section{Frequency mixing}
\label{sec:phase}

Shifting a signal in frequency by mixing it with a local oscillator signal, as in a typical radio receiver, also changes the phase of the signal.  To find a possible peak in the original signal, it is necessary to search through the possible values of this phase to find the maximum peak height.  Here we illustrate this effect, and show that the phase-searched maximum signal can be represented by an analytic envelope of the signal.

\subsection{Origin of the phase ambiguity}
\label{sec:phase-ambiguity}

The mixing stage in a radio receiver can be represented as a multiplication between the original radio frequency signal $f\subRF(t)$ and a local oscillator signal $f\subLO(t)$, equivalent to a convolution $F\subRF(\nu) * F\subLO(\nu)$ in the frequency domain.  The local oscillator and its Fourier transform $F\subLO(\nu)$ are given by
\begin{equation}
 \begin{aligned}
  f\subLO(t) &= \cos( 2 \pi \, \nu\subLO \, t \, + \, \phi ) \\
  F\subLO(\nu) &= \frac{1}{2} \, e^{i \phi} \, \delta( \nu - \nu\subLO ) + \frac{1}{2} \, e^{-i \phi} \, \delta( \nu + \nu\subLO )
 \end{aligned}
 \label{eqn:lo}
\end{equation}
where $\phi$ describes the phase of the local oscillator relative to the signal.  Since the delay between the local oscillator and the radio frequency signal is arbitrary, this phase is typically unknown.  After subsequent bandpass filtering to obtain the low-frequency component, the resulting intermediate frequency signal $f\subIF(t)$ is described, in the frequency domain, as:
\begin{align}
 F\subIF(\nu) &= \frac{1}{2} \, e^{i \phi} \, F\subRF( \nu - \nu\subLO ) + \frac{1}{2} \, e^{-i \phi} \, F\subRF( \nu + \nu\subLO ) & \mbox{$|\nu| < \nu\subLO$} .
\end{align}
For a low-side local oscillator (i.e. $\nu\subLO$ is less than all frequencies $\nu$ present in the radio frequency signal), this becomes
\begin{equation}
 F\subIF(\nu) = \frac{1}{2} \, e^{-i \phi \, \sign(\nu)}
  \begin{cases}
   F\subRF( \nu + \nu\subLO ) & 0 < \nu < \nu\subLO \\
   F\subRF( \nu - \nu\subLO ) & -\nu\subLO < \nu < 0
  \end{cases}
 \label{eqn:if_final}
\end{equation}
showing that the intermediate frequency signal depends on the phase factor $e^{-i \phi \, \sign(\nu)}$.  The expression for a high-side local oscillator is similar, but with the sign of $\phi$ reversed; additional mixing stages add additional phases $\phi_1$, $\phi_2$, etc.  If the signal is mixed separately with two local oscillators out of phase by $\pi/2$, as in I/Q demodulation, the resulting intermediate frequency signals have phases $\phi_I$ and \mbox{$\phi_Q = \phi_I + \pi/2$}.  In all these cases, if the phase of the local oscillator is unknown, the effect is the same: the intermediate frequency signal includes an unknown phase $\phi$ in the range $0 \to 2\pi$.  Consequently, from equation~\ref{eqn:if_final}, it is not possible to completely determine the radio frequency signal $F\subRF(\nu)$ from the intermediate frequency signal $F\subIF(\nu)$.

\subsection{Phase searching and the signal envelope}
\label{sec:phase-searching}

\begin{figure}
 \centering
 \includegraphics[width=\linewidth]{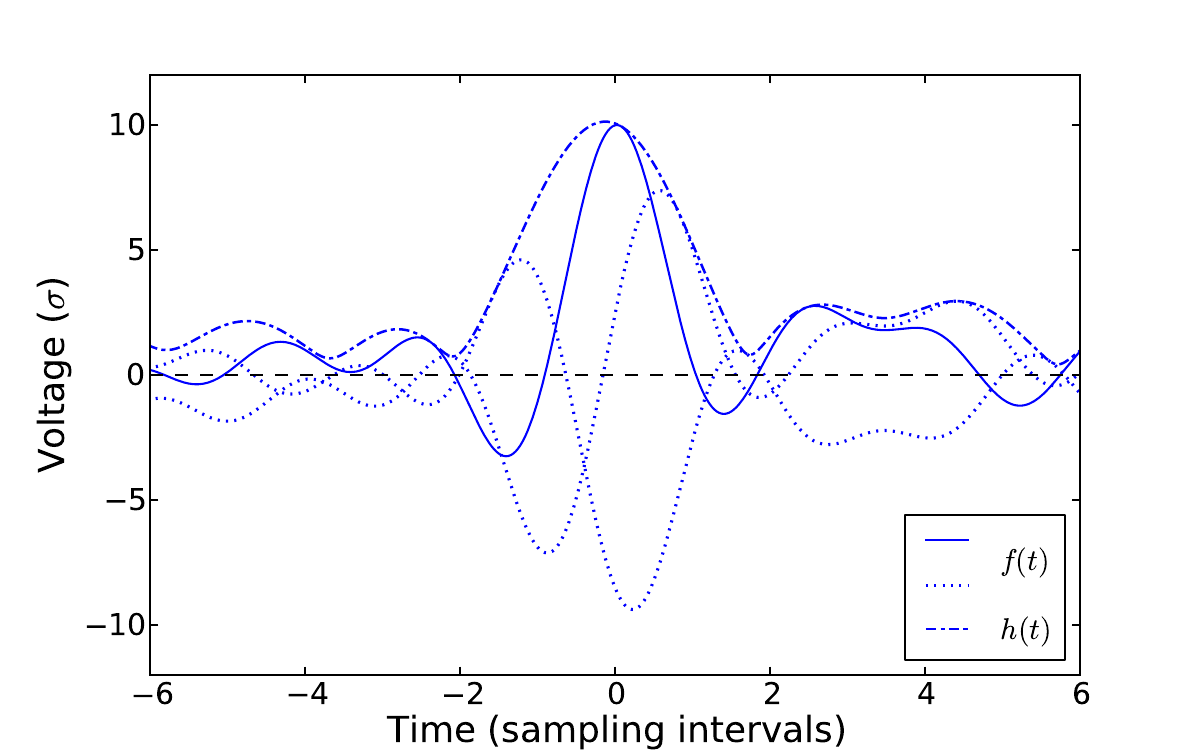}
 \caption{A coherent, zero-phase, flat-spectrum pulse (i.e. a $\sinc$ function) against a background of Gaussian random noise with the same spectrum (solid).  Adding a variable phase $\phi$ to the signal, as per equation~\ref{eqn:if_final}, results in a range of possible pulse shapes, some of which are shown (dotted), with a decreased peak amplitude.  These represent different possibilities for the signal $f(t)$ from section~\ref{sec:phase-searching}.  The signal envelope $h(t)$ (dash-dotted), from equation~\ref{eqn:h(t)}, recovers the full amplitude of the pulse.}
 \label{fig:mixing}
\end{figure}

The effect in the time domain of introducing a variable phase $\phi$ can be to decrease the peak amplitude of the pulse, as illustrated in figure~\ref{fig:mixing}.  As $\phi$ is unknown, it is not possible to unambiguously reconstruct the original signal.  However, given a post-mixing signal $f(t) = f\subIF(t)$, it is possible to reconstruct the different possible states of the original signal for different values of $\phi$.  An excursion beyond some threshold in the original signal must appear as a peak of at least that height in one or more of these reconstructed signals.  We define $h(t,\theta)$ as the reconstructed form of $f(t)$ with maximum peak height; in terms of their respective Fourier transforms, $H(\nu,\theta)$ and $F(\nu)$, it is
 \begin{equation}
  H(\nu, \theta) = e^{i \theta \, \sign(\nu)} F(\nu)
 \end{equation}
where $\theta$ is chosen to maximise the peak in $h(t,\theta)$.  For $\theta = \phi$ this will restore the phase of the original signal (cf. equation~\ref{eqn:if_final}), but this is not generally the case.  We can re-express the reconstructed signal as
 \begin{equation}
  H(\nu, \theta) = F(\nu) \cos\theta + G(\nu) \sin\theta \label{eqn:H(nu)}
 \end{equation}
where
 \begin{equation}
  G(\nu) = i \, \sign(\nu) \, F(\nu) . \label{eqn:G(nu)}
 \end{equation}
Equivalently to equation~\ref{eqn:H(nu)}, in the time domain,
 \begin{equation}
  h(t, \theta) = f(t) \cos\theta + g(t) \sin\theta . \label{eqn:h(t)_early}
 \end{equation}
Since $\theta$ is chosen to maximise $h(t,\theta)$, we know that
 \begin{equation}
  \frac{\partial}{\partial\theta} h(t, \theta) = 0 ,
 \end{equation}
hence
 \begin{equation}
  \frac{g(t)}{f(t)} = \tan\theta
 \end{equation}
and $h(t,\theta)$ from equation~\ref{eqn:h(t)_early} can be rewritten without $\theta$-dependence as
 \begin{equation}
  h(t) = \sqrt{f(t)^2 + g(t)^2} \label{eqn:h(t)} ,
 \end{equation}
where the choice of the positive square root reflects the choice of a positive peak in $h(t)$.  We can express $g(t)$ by using the convolution theorem to find the time-domain equivalent to equation~\ref{eqn:G(nu)},
 \begin{equation}
  g(t) = \frac{1}{\pi \, t} * f(t) , \label{eqn:g(t)}
 \end{equation}
which is the Hilbert transform of $f(t)$.  This makes $h(t)$ the same as the signal envelope defined and extensively discussed by \citet{longuet-higgins1957}; here we will consider only its excursion rate, in sections~\ref{sec:exrate_env} and~\ref{sec:exrate_interpenv}.

\section{Excursion rates}
\label{sec:exrate}

The significance of an event detected in a stochastic signal is dependent upon the expected rate of false detections, given the characteristics of the signal.  Here we consider the case of pulse-like events, involving an excursion beyond some high threshold, in a Gaussian random signal with an arbitrary spectrum, and find the expected rate of such excursions.

We give the excursion rate for a raw digital signal (i.e. after sampling, but before further digital processing) in section~\ref{sec:exrate_rawsig}, and after perfect interpolation (i.e. for a continuous signal) in section~\ref{sec:exrate_interpsig}.  The excursion rate for the signal envelope (as defined in section~\ref{sec:phase-searching}) is given in section~\ref{sec:exrate_env}; and a new result for the interpolated signal envelope, based on the previous results, is derived in section~\ref{sec:exrate_interpenv}.

The excursion rates given here are one-sided, beyond a positive threshold only, following the convention of \citet{rice1944,rice1945}.  For a Gaussian signal, the two-sided excursion rates will be exactly twice these values.  The signal envelope is always positive, so for it the distinction is unimportant.

\subsection{Raw signal}
\label{sec:exrate_rawsig}

Immediately after sampling, a signal such as the measured voltage in a radio receiver consists of a series of Gaussian samples $f(t_i)$ with probability density
 \begin{equation}
  P(f) = \frac{ 1 }{ \sigma \sqrt{2 \pi} } \, \expfrac[-]{f^2\!}{2\sigma^2}
  \label{eqn:gauss_dist}
 \end{equation}
where $\sigma$ is the standard deviation or root mean square (RMS) signal amplitude, and the mean is assumed to be zero.  An excursion in the signal consists of one or more consecutive samples above some threshold $\fthr$.  If $\fthr$ is sufficiently high that it is rare for two consecutive samples to exceed it, the probability that a particular sample exceeds the threshold gives us the expected rate of such excursions,
 \begin{align}
  \textrm{excursion rate} &= \int_{\fthr}^{\infty} \! df \, P(f) \nonumber \\
   &= \frac{1}{2} - \frac{1}{2} \, \erf\!\left( \frac{\fthr}{\sigma\sqrt{2}} \right) \label{eqn:exrate_rawsig}
 \end{align}
in units of the sampling rate, with $\erf$ being the standard Gauss error function.  However, if $\fthr$ is low, or if the signal samples are strongly correlated in time --- as for an oversampled signal, or after interpolation --- a single excursion will typically involve a number of consecutive samples, so this approach is insufficient.

\subsection{Interpolated signal}
\label{sec:exrate_interpsig}

After interpolation to produce a continuous signal, the excursion rate is found by means of Rice's formula, which gives the rate of upward crossings --- and, hence, distinct excursions --- over a threshold $\fthr$ for a signal $f = f(t)$, as
 \begin{equation}
  \textrm{excursion rate} = \int_0^\infty \! df' \, f' \, P(\fthr, f')
  \label{eqn:rice}
 \end{equation}
where $f'$ is the time derivative of $f$, and $P(f, f')$ is their joint probability distribution.  An early form of this result appears in \citet[eqn 3.3-5]{rice1945}; refer to \citet{rainal1988} for a detailed discussion.  $P(f, f')$ can be rewritten as
 \begin{equation}
  P(f, f') = P(f) \, P(f'; f)
 \end{equation}
where $P(f)$ is the distribution of $f$ (from equation~\ref{eqn:gauss_dist}) and $P(f'; f)$ is the distribution of $f'$ conditional upon the value of $f$.  To find this latter distribution, we consider points at closely separated times $t$ and $t+dt$.  If $f(t)$ is a correlated Gaussian signal, then the corresponding values $f$ and $f+df$ follow a joint normal distribution (see e.g. \citet[ch. 4]{kenney1939b}): given the value of $f$, $f+df$ is normally distributed with expected value and variance
 \begin{equation}
  \begin{aligned}
   \expec{f+df} &= f \, \rho(dt) \\
   \var{f+df} &= \sigma^2 \left( 1 - \rho(dt)^2 \right)
  \end{aligned}
  \label{eqn:pt2_dist}
 \end{equation}
where $\rho(dt)$ is the value of the autocorrelation function (Fourier transform of the power spectrum, normalised to $\rho(0) = 1$) at separation $dt$.  The slope $f' = df/dt$ is then also normally distributed, with expected value and variance
 \begin{equation}
  \begin{aligned}
   \expec{f'} &= \frac{ \expec{f+df} - f }{ dt } \\
   \var{f'} &= \frac{ \var{f+df} } { dt^2 } .
  \end{aligned}
 \end{equation}
Taking the limit $dt \to 0$ and the expressions from equation~\ref{eqn:pt2_dist}, and noting that $\rho'(0) = 0$, we obtain
 \begin{equation}
  \begin{aligned}
   \expec{f'} &= 0 \\
   \var{f'} &= - \sigma^2 \, \rho''(0)
  \end{aligned}
 \end{equation}
which represents the distribution
 \begin{align}
  P(f'; f) &= P(f') & \mbox{irrespective of $f$} \nonumber \\
   &= \frac{ 1 }{ \fprimestd \sqrt{2 \pi} } \, \expfrac[-]{f'^2\!}{2\fprimestd^2} \label{eqn:slope_dist}
 \end{align}
where
 \begin{align}
  \fprimestd &= \sqrt{\var{f'}} \nonumber \\
   &= \sigma \sqrt{ -\rho''(0) } . \label{eqn:sigma_m}
 \end{align}
With the expressions for $P(f)$ and $P(f'; f)$ from equations~\ref{eqn:gauss_dist} and~\ref{eqn:slope_dist} respectively, equation~\ref{eqn:rice} gives us
 \begin{equation}
  \textrm{excursion rate} = \frac{1}{2\pi} \sqrt{ - \rho''(0) } \, \expfrac[-]{\fthr^2}{2\sigma^2}
  \label{eqn:exrate_interpsig}
 \end{equation}
as found by \citet[eqns 3.3-11 \& 3.6-11]{rice1945}.  For discrete data, as in section~\ref{sec:bedlam} of this work, $\rho(t)$ can be obtained numerically from the measured power spectrum.

\subsubsection{White noise}
\label{sec:exrate_interpsig_white}

When the power spectrum is known analytically, equation~\ref{eqn:exrate_interpsig} can be simplified.  Consider the special case of white noise, with constant power spectral intensity from frequency $\nu_a$ to $\nu_b$, i.e.
 \begin{equation}
  I(\nu) = 
  \begin{cases}
   1 & \nu_a < |\nu| < \nu_b \\
   0 & \mbox{otherwise} .
  \end{cases}
 \end{equation}
This function has the Fourier transform
 \begin{equation}
  \mathcal{F}\{I\}(t) = 2 \, \nu_b \, \sinc(2 \, \nu_b \, t) - 2 \, \nu_a \, \sinc(2 \, \nu_a \, t)
 \end{equation}
where $\sinc(x) = \sin(\pi x) / (\pi x)$.  Normalising this, we obtain the autocorrelation function
 \begin{equation}
  \rho(t) = \frac{ \nu_b \, \sinc(2 \, \nu_b \, t) - \nu_a \, \sinc(2 \, \nu_a \, t) }{\nu_b - \nu_a} . \label{eqn:white_rho}
 \end{equation}
The second derivative of this function at $t = 0$ is
 \begin{equation}
  \rho''(0) = - \frac{4}{3} \, \pi^2 \, \frac{ \nu_a^3 - \nu_b^3 }{ \nu_a - \nu_b } , \label{eqn:rho''(0)}
 \end{equation}
which we insert into equation~\ref{eqn:exrate_interpsig} to obtain
 \begin{equation}
  \textrm{excursion rate} = \sqrt{ \frac{1}{3} \frac{ \nu_a^3 - \nu_b^3 }{ \nu_a - \nu_b } } \, \expfrac[-]{\fthr^2}{2\sigma^2}, \label{eqn:exrate_interpsig_white}
 \end{equation}
matching the result found for this case by \citet[eqns 3.3-12 \& 3.6-11]{rice1945}.

\subsection{Signal envelope}
\label{sec:exrate_env}

The signal envelope $h(t)$, per equation~\ref{eqn:h(t)}, is a function of the signal $f(t)$ and its Hilbert transform $g(t)$.  We take both $f = f(t)$ and $g = g(t)$ to be normally distributed with standard deviation $\sigma$.  The distribution of $h = h(t)$ depends on the degree of correlation between them.  Given the value of $f(t)$, the expected value of the signal at other times is
 \begin{equation}
  \expec{f(t-\tau)} = f(t) \, \rho(\tau) \label{eqn:E[f(t)]}
 \end{equation}
where $\rho(\tau)$ is the autocorrelation function as in section~\ref{sec:exrate_interpsig}.  The expected value of $g$, from equation~\ref{eqn:g(t)}, is then
 \begin{align}
  \expec{g} 
   &= \int\! d\tau \, \frac{1}{\pi \tau} \, \expec{f(t-\tau)} \nonumber \\
   &= \int\! d\tau \, \frac{1}{\pi \tau} \, f(t) \, \rho(\tau) \nonumber \\
   &= 0 & \mbox{as $\displaystyle\frac{1}{\pi \tau}$ is odd, and $\rho(\tau)$ is even.} \label{eqn:fg_corr}
 \end{align}
Therefore there is no correlation between $f$ and $g$; and, given that they are normally distributed, equation~\ref{eqn:h(t)} implies that $h$ follows a Rayleigh distribution
 \begin{equation}
  P(h) = \frac{h}{\sigma^2} \, \expfrac[-]{h^2\!}{2 \sigma^2} \label{eqn:chi}
 \end{equation}
as found by \citet{longuet-higgins1957}.  If consecutive values $h(t_i)$ are not significantly correlated, and we choose a threshold $\hthr$ sufficiently high that consecutive values are unlikely to exceed it (as in section~\ref{sec:exrate_rawsig}), this gives us the excursion rate of the signal envelope:
 \begin{align}
  \textrm{excursion rate} &= \int_{\hthr}^{\infty} \! dh \, P(h) \nonumber \\
   &= \expfrac[-]{\hthr^2}{2 \sigma^2} \label{eqn:exrate_env}
 \end{align}
in units of the sampling rate.

\subsection{Interpolated signal envelope}
\label{sec:exrate_interpenv}

The excursion rate for the interpolated signal envelope $h$ has not been addressed in previous literature.  To find it, we apply the approach from section~\ref{sec:exrate_interpsig}, which requires us to find the distribution of its time derivative $h'$.  From equation~\ref{eqn:h(t)},
 \begin{equation}
  h' = \frac{1}{h} \left( f \, f' + g \, g' \right) . \label{eqn:s}
 \end{equation}
The distribution of this value depends on the correlation between its components.  We know from section~\ref{sec:exrate_interpsig} that $f'$ is normally distributed with mean $0$ and variance $\fprimestd^2$, and uncorrelated with $f$; the same applies to $g'$ with respect to $g$.  We also know from section~\ref{sec:exrate_env} that $f$ and $g$ are uncorrelated.  We now consider the correlation between $f$ and $g'$, the latter of which is found as
 \begin{align}
  g'(t) &= \frac{d}{d t} \left( \frac{1}{\pi t} * f(t) \right) & \mbox{from equation~\ref{eqn:g(t)}} \nonumber \\
   &= - \frac{1}{\pi t^2} * f(t) .
 \end{align}
Hence the conditional expected value of $g'$, given $f$, is
 \begin{align}
  \expec{g'} &= - \int\! d\tau \, \frac{1}{\pi \tau^2} \, \expec{f(t - \tau)} \nonumber \\
   &= - \int\! d\tau \, \frac{1}{\pi \tau^2} \, f(t) \, \rho(\tau) & \mbox{from equation~\ref{eqn:E[f(t)]}} \nonumber \\
   &= \fgcorr \, \frac{\fprimestd}{\sigma} \, f \label{eqn:E[g'(t)]}
 \end{align}
where
 \begin{align}
  \fgcorr &= - \frac{\sigma}{\fprimestd} \int\! d\tau \, \frac{1}{\pi \tau^2} \, \rho(\tau) \nonumber \\
   &= \frac{-1}{\sqrt{-\rho''(0)}} \int\! d\tau \, \frac{1}{\pi \tau^2} \, \rho(\tau) & \mbox{from equation~\ref{eqn:sigma_m}} \label{eqn:rho_m}
 \end{align}
is the coefficient of correlation between $f$ and $g'$.  The variance of $g'$ is then
 \begin{equation}
  \var{g'} = \fprimestd^2 \left( 1 - \fgcorr^2 \right) \label{eqn:Var[g'(t)]}
 \end{equation}
conditional upon the value of $f$.

We consider the correlation between $g$ and $f'$ in a similar manner.  First we note, from equation~\ref{eqn:g(t)}, that we can obtain $f$ via the inverse Hilbert transform
 \begin{equation}
  f(t) = - \frac{1}{\pi t} * g(t) .
 \end{equation}
Hence
 \begin{align}
  f'(t) &= \frac{d}{d t} \left( - \frac{1}{\pi t} * g(t) \right) \nonumber \\
   &= \frac{1}{\pi t^2} * g(t)
 \end{align}
and the conditional expected value, given $g$, is
 \begin{equation}
  \expec{f'} = - \fgcorr \frac{\fprimestd}{\sigma} \, g . \label{eqn:E[f'(t)]}
 \end{equation}
The conditional variance, $\var{f'}$, is identical to $\var{g'}$ in equation~\ref{eqn:Var[g'(t)]}, as it is insensitive to the sign of $\fgcorr$.

We know that $h'$, given $f$ and $g$, is the sum of two independent normal variables (corresponding to the terms in equation~\ref{eqn:s}), and hence normally distributed itself.  We find its expected value and variance:
 \begin{align}
  \expec{h'} &= \frac{1}{h} \left( f \, \expec{f'} + g \, \expec{g'} \right) \nonumber \\
   &= 0 & \mbox{from equations~\ref{eqn:E[g'(t)]} and \ref{eqn:E[f'(t)]}} \\
  \var{h'} &= \frac{1}{h^2} \left( f^2 \, \var{f'} + g^2 \, \var{g'} \right) \nonumber \\
   &= \frac{f^2 + g^2}{h^2} \, \var{f'} & \mbox{as $\var{f'} = \var{g'}$} \nonumber \\
   &= \fprimestd^2 \left( 1 - \fgcorr^2 \right) & \mbox{from equations~\ref{eqn:h(t)} and~\ref{eqn:Var[g'(t)]}} . \label{eqn:var_s}
 \end{align}
The distribution of $h'$ is thus
 \begin{align}
  P(h'; h) &= P(h') & \mbox{irrespective of $h$} \nonumber \\
   &= \frac{1}{\hprimestd \sqrt{2\pi}} \expfrac[-]{h'^2\!}{2 \hprimestd^2}
 \end{align}
where
 \begin{align}
  \hprimestd &= \sqrt{ \var{h'} } \nonumber \\
   &= \sigma \sqrt{-\rho''(0)} \sqrt{1 - \fgcorr^2} & \mbox{from equations~\ref{eqn:sigma_m} and \ref{eqn:var_s}} . \label{eqn:phasedslope_stdev}
 \end{align}
With this distribution, and the distribution $P(h)$ from equation~\ref{eqn:chi}, we can determine the excursion rate beyond a threshold $\hthr$ from equation~\ref{eqn:rice}, finding that
 \begin{equation}
  \textrm{excursion rate} = \frac{1}{\sqrt{2\pi}} \, \sqrt{ -\rho''(0) - \left( \int \! d\tau \, \frac{1}{\pi \tau^2} \, \rho(\tau) \right)^2 } \, \frac{\hthr}{\sigma} \, \expfrac[-]{\hthr^2}{2\sigma^2} . \label{eqn:exrate_interpenv}
 \end{equation}
If the integral in this expression is to be evaluated with numerical data, with values of $\rho(t)$ calculated for discrete time intervals, note that the analytic result
 \begin{equation}
  \int_{-\varepsilon}^{\varepsilon} \! d\tau \, \frac{1}{\pi \tau^2} \, \rho(\tau) \simeq - \frac{2}{\pi\varepsilon} \, \rho(0)
 \end{equation}
should be used for the interval $-\varepsilon < \tau < \varepsilon$ to avoid the singular behaviour at $\tau = 0$.

\subsubsection{White noise}

Here we consider white noise from frequency $\nu_a$ to $\nu_b$ as in section~\ref{sec:exrate_interpsig_white}, and derive the excursion rate for the signal envelope in this case.  The autocorrelation function for white noise is given by equation~\ref{eqn:white_rho}, which allows us to evaluate the integral in equation~\ref{eqn:exrate_interpenv} as
 \begin{equation}
  \int \! d\tau \, \frac{1}{\pi \tau^2} \, \rho(\tau) = -\pi (\nu_a + \nu_b) .
 \end{equation}
Also incorporating $\rho''(0)$ from equation~\ref{eqn:rho''(0)}, we can evaluate the excursion rate from equation~\ref{eqn:exrate_interpenv} to be
 \begin{equation}
  \textrm{excursion rate} = \sqrt{\frac{\pi}{6}} \, | \nu_a - \nu_b | \, \frac{\hthr}{\sigma} \, \expfrac[-]{\hthr^2}{2\sigma^2} .
  \label{eqn:exrate_interpenv_white}
 \end{equation}

\section{Monte Carlo simulation}
\label{sec:mc}

The results from section~\ref{sec:exrate} can be tested through Monte Carlo simulation: generating random noise data and comparing the distribution of peak heights (for the signal and its envelope, with and without interpolation) with the analytic predictions.  This also provides a helpful illustration of the effects of interpolation and the phase search represented by the signal envelope on the excursion rate.

\subsection{Uncorrelated noise}
\label{sec:mc_white}

The simplest case to test through simulation is that of uncorrelated noise.  We generated \bignum{100}{000} arrays, each containing \bignum{4}{096} independent normally-distributed random values.  For each array, we performed the following steps:
\begin{enumerate}
 \item Find the maximum value in the array.
 \item Perform interpolation on the array, and find the maximum interpolated value. \label{it:interp}
 \item Take the norm of the array and its Hilbert transform to find the signal envelope, as in equation~\ref{eqn:h(t)}.  Find the maximum value in the resulting array. \label{it:phase}
 \item Repeat step~\ref{it:phase} with an interpolated version of the array, and find the maximum value. \label{it:both}
\end{enumerate}
The interpolation in steps~\ref{it:interp} and~\ref{it:both} was 32-fold, with 31 interpolated values between each pair of raw values.  This level of precision was chosen as it resulted in maximum reconstructed values within 0.1\% of those resulting from finer searches, while being relatively inexpensive to compute.

The number of excursions beyond some threshold, under each of the conditions above (raw values; interpolation; signal envelope; interpolated envelope), can be considered to be the number of arrays containing a peak height above that threshold.  This leads to an underestimate of the excursion rate for low thresholds, for which multiple excursions within a single \bignum{4}{096}-value array are common; but it is accurate for high thresholds, for which this is rare.

\begin{figure}
 \centering
 \includegraphics[width=\linewidth]{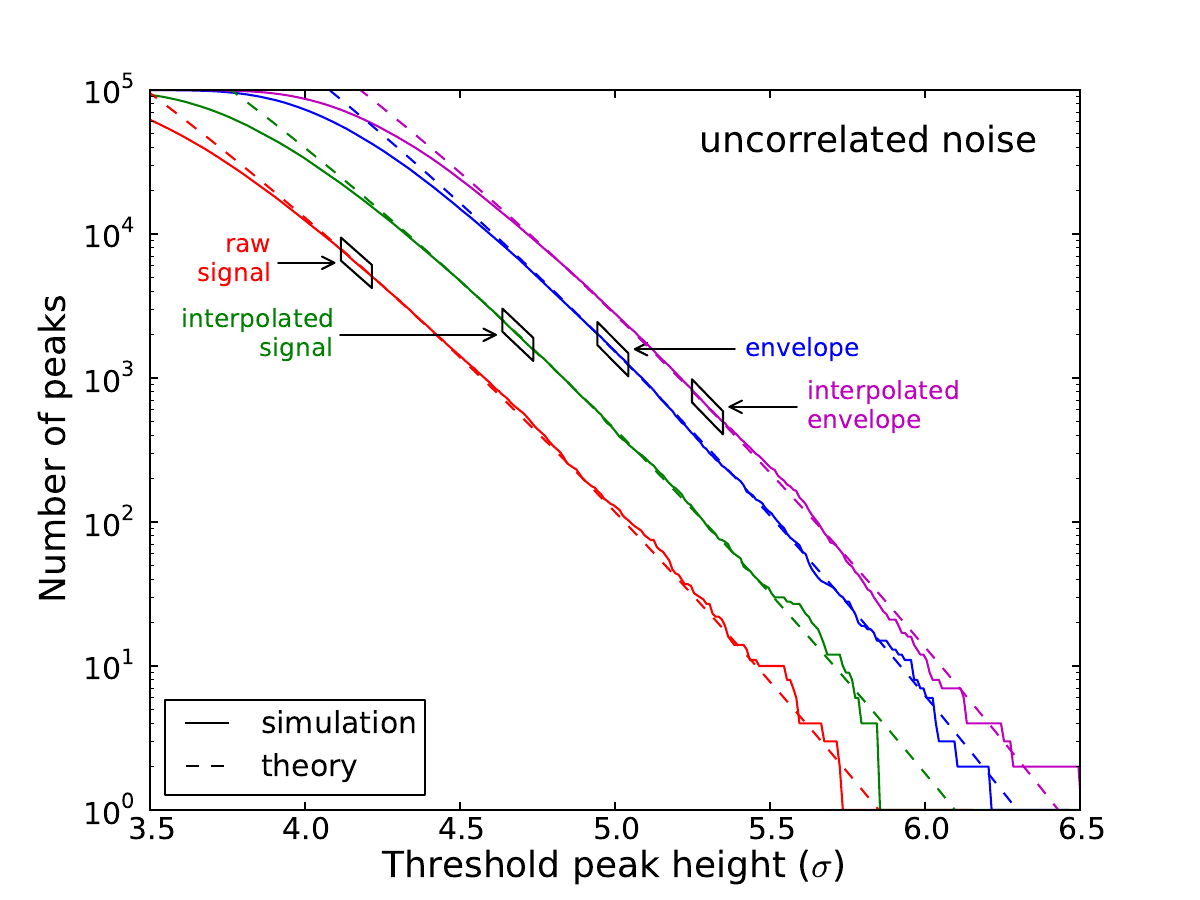}
 \caption{Cumulative histograms (solid) of the peak height in an array for white-spectrum Monte Carlo data, for the signal and its envelope, with and without interpolation, compared with analytic predictions (dashed).  The divergence at low thresholds is due to arrays containing multiple peaks, which are counted only once.  The divergence at high thresholds is due to small-number statistics.}
 \label{fig:hist_mc_white}
\end{figure}

As the values are uncorrelated, the expected excursion rates at high thresholds for the uninterpolated data are given by equation~\ref{eqn:exrate_rawsig} for the raw signal and equation~\ref{eqn:exrate_env} for the signal envelope.  Uncorrelated values also result in a white spectrum, with a constant spectral intensity up to an implied maximum frequency of 0.5 cycles per sampling interval, allowing the use of equations~\ref{eqn:exrate_interpsig_white} and~\ref{eqn:exrate_interpenv_white} to predict the excursion rates for interpolated data, for the signal and its envelope respectively.  In figure~\ref{fig:hist_mc_white} we compare these excursion rates, in terms of the total expected number of excursions, with cumulative histograms of the peak height in an array, finding them in good agreement.

\subsection{Correlated noise}
\label{sec:mc_gauss}

To test the dependence of the excursion rate on the autocorrelation of the noise, we performed another Monte Carlo simulation with a different spectrum.  Arrays of random values were generated with a Gaussian power spectrum
 \begin{align}
  I(\nu) & \propto \expfrac[-]{\nu^2\!}{2\sigma_\nu^2} & |\nu| < \nu_s / 2
   \label{eqn:gauss_spect}
 \end{align}
where $\sigma_\nu = \nu_s / 5$ is defined in terms of the sampling rate $\nu_s$.  Peak heights for the signal and its envelope, with and without interpolation, were then found through the same procedure as in section~\ref{sec:mc_white}.

As the values in this case are correlated, the excursion rates for interpolated data must be found with equation~\ref{eqn:exrate_interpsig} for the signal and equation~\ref{eqn:exrate_interpenv} for its envelope, in terms of the autocorrelation function $\rho(t)$.  This function can be derived analytically from the power spectrum in equation~\ref{eqn:gauss_spect}, but was instead obtained numerically in order to validate the procedure to be used in section~\ref{sec:bedlam}.  Unlike the uncorrelated data in section~\ref{sec:mc_white}, excursion rates for correlated uninterpolated data are not expected to be given exactly by equations~\ref{eqn:exrate_rawsig} and~\ref{eqn:exrate_env}.

\begin{figure}
 \centering
 \includegraphics[width=\linewidth]{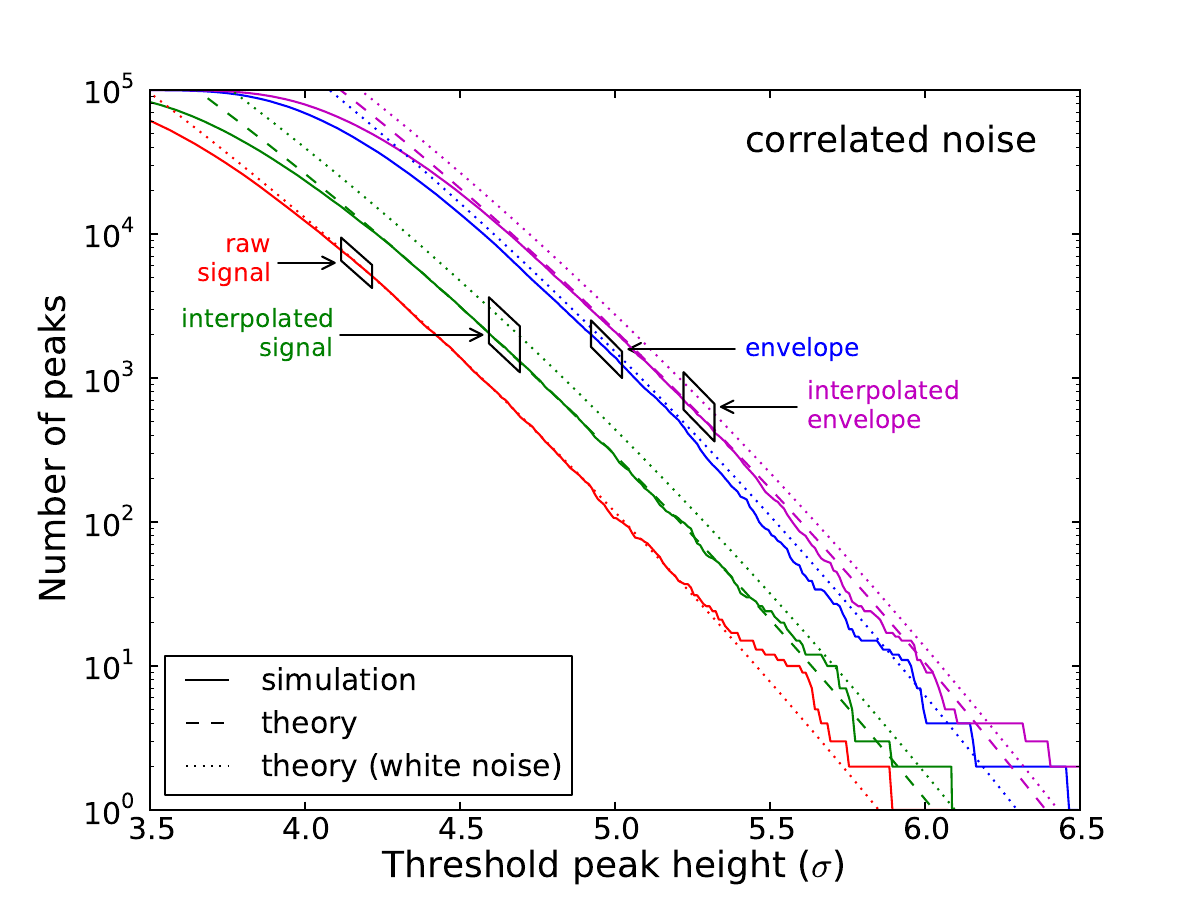}
 \caption{Cumulative histograms (solid) of the peak height in an array for Gaussian-spectrum Monte Carlo data, for the signal and its envelope, with and without interpolation, compared with analytic predictions for the interpolated case (dashed).  (We have not derived analytic predictions for the uninterpolated case for correlated noise.)  Divergence at low and high thresholds, as in figure~\ref{fig:hist_mc_white}, is due to under-counting of peaks and small-number statistics respectively.  Analytic predictions for the case of uncorrelated noise (dotted) are shown for comparison.}
 \label{fig:hist_mc_gauss}
\end{figure}

Cumulative histograms of peak heights are compared with these predictions in figure~\ref{fig:hist_mc_gauss}.  The histograms for interpolated data agree closely with the predictions, significantly lower than the excursion rates expected for uncorrelated noise.

\section{Experimental test with the Parkes Bedlam backend}
\label{sec:bedlam}

Bedlam is a digital backend for the Parkes radio telescope, designed and built to perform transient pulse detection for the LUNASKA project~\citep{bray2010}.  We performed an experiment in which Bedlam processed data from a noise source, the results of which allow us to test the performance of the hardware and the results from section~\ref{sec:exrate}.

\subsection{The Parkes Bedlam backend}
\label{sec:sigpath}

In normal operation, Bedlam is connected to the Parkes 21 cm multibeam receiver~\citep{staveley-smith1996}.  It has eight inputs, typically receiving the voltage signal from both polarisations for four of this receiver's thirteen beams.  The receiver amplifies the signal and downconverts it to a lower frequency (see section~\ref{sec:phase-ambiguity}); it then enters Bedlam where it is digitised by analog-to-digital converters (ADCs) at 1,024 Msample/s (corresponding to 512 MHz of bandwidth) and 8 bits of precision, and subjected to an adjustable dedispersion filter to compensate for ionospheric dispersion.  Buffers of the raw and dedispersed signals are kept, with an adjustable size up to 8 $\mu$s.

The dedispersed data are then interpolated.  Full reconstruction of a Nyquist-sampled signal is achieved by convolving with a $\sinc$ function; interpolation here is performed as a convolution with a six-point approximation to a $\sinc$ function (at $\pm 0.5$, $\pm 1.5$ and $\pm 2.5$ sampling intervals), producing a single interpolated point between every pair of original samples.  If a value is found over an adjustable threshold, in either the original or interpolated values, it triggers the permanent storage of both raw and dedispersed buffers for all input channels.  Anticoincidence logic can be applied between channels to prevent triggering from simultaneous above-threshold values in multiple channels, but was inactive for this experiment.

\begin{figure}
 \centering
 \includegraphics[width=\linewidth]{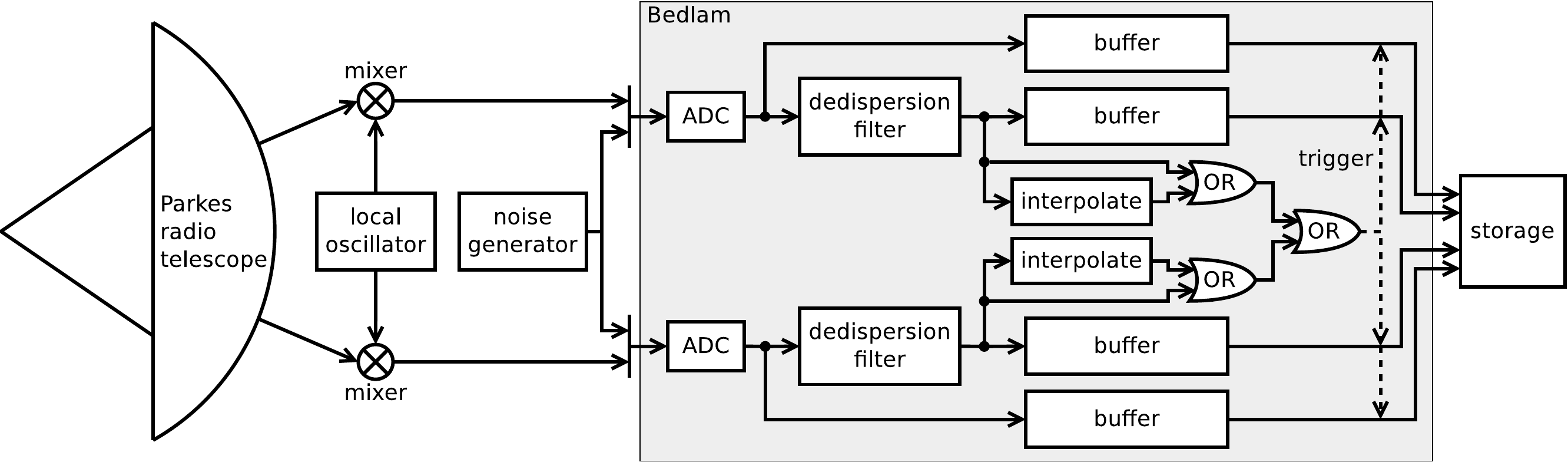}
 \caption{Signal path for two active channels in the Parkes Bedlam backend (see section~\ref{sec:sigpath} for details).  In normal operation (outer signal paths) it receives input from the Parkes radio telescope; in this experiment, it was instead connected to a noise generator (centre).  A trigger condition (dashed) causes the contents of the buffers to be stored.}
 \label{fig:sigpath}
\end{figure}

In this experiment, all eight channels were connected to the same noise source.  However, thresholds were set such that only two of these were able to trigger buffer storage.  The signal path for these channels is shown in figure~\ref{fig:sigpath}.

\subsection{Experiment}
\label{sec:experiment}

The Bedlam backend was operated with pure noise input for 478 hours, starting on 23 September 2010.  The noise source was a reverse-biased diode operating at its breakdown voltage, with a passive eight-way power-splitter to connect it to all eight input channels.  In this time, there were \bignum{902}{115} peaks in the noise signal sufficient to trigger storage of the buffered data, with the buffer size set to 4 $\mu$s (\bignum{4}{096} samples).  During the 51 ms period taken to store a single set of buffers of this length, the system is unable to respond to any further triggers; this dead time amounts to 2.7\% of the experimental period.  The dedispersion filter was set to its minimal setting, equivalent to $\sim 2$ ns differential delay across the band of the noise.

The RMS noise amplitude was typically $\sim 12$ analog-to-digital units (ADU), compared to a maximum of +127 ADU for an 8-bit signed integer.  The temperature of the noise source was not controlled, so the gain varied over the course of the experiment, resulting in $\sim 5\%$ slow variation in the RMS.  More dramatic fluctuations in the RMS occurred during an 8-hour period partway through the experiment, possibly due to a problem with the power supply, so this interval was excluded.  After also excluding dead time, the remaining effective duration of the experiment is 458 hours.

\begin{figure}
 \centering
 \includegraphics[width=\linewidth]{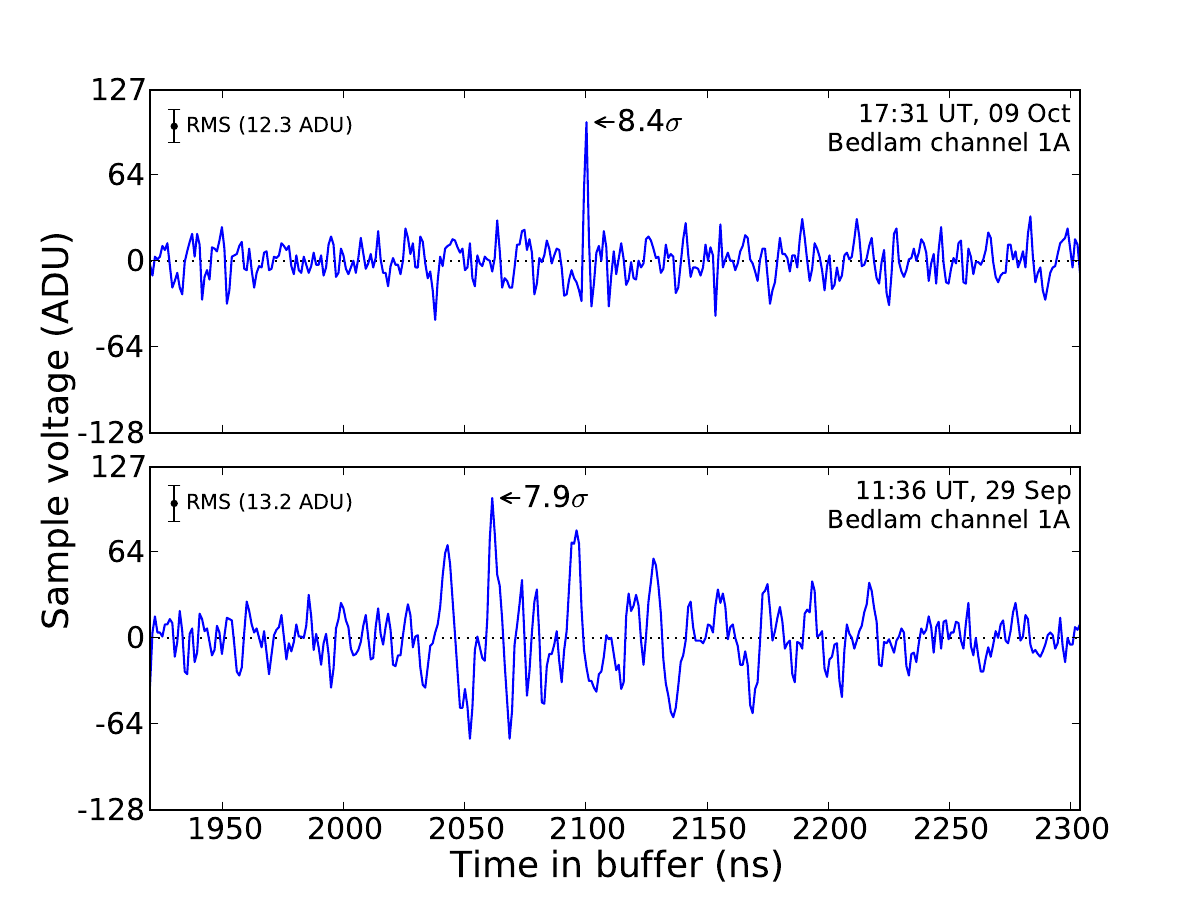}
 \caption{The two most significant peaks, after dedispersion, recorded over the course of the experiment.  The upper peak displays the narrow profile typical of rare fluctuations in Gaussian noise.  The lower peak includes an extended elevation in signal power, possibly indicative of an instrumental problem, and is one of two such peaks excluded from the analysis (see text).  In both cases, similar peaks are seen in the other channels.}
 \label{fig:badpulse}
\end{figure}

Typical stored buffers consisted mostly of pure noise, with a single extreme value that exceeded the trigger threshold in the centre of the buffer, as expected from Gaussian noise with a relatively wide band, as in figure~\ref{fig:badpulse} (top).  However, there were two atypical events which exhibited increased signal power for an extended period, over 100 ns, as in figure~\ref{fig:badpulse} (bottom).  These events are not consistent with pure Gaussian noise; their origin is unknown, but more such events were found during the period excluded due to RMS fluctuations, which suggests that they may be related phenomena.  These two events were excluded from further analysis.

\begin{figure}
 \centering
 \includegraphics[width=\linewidth]{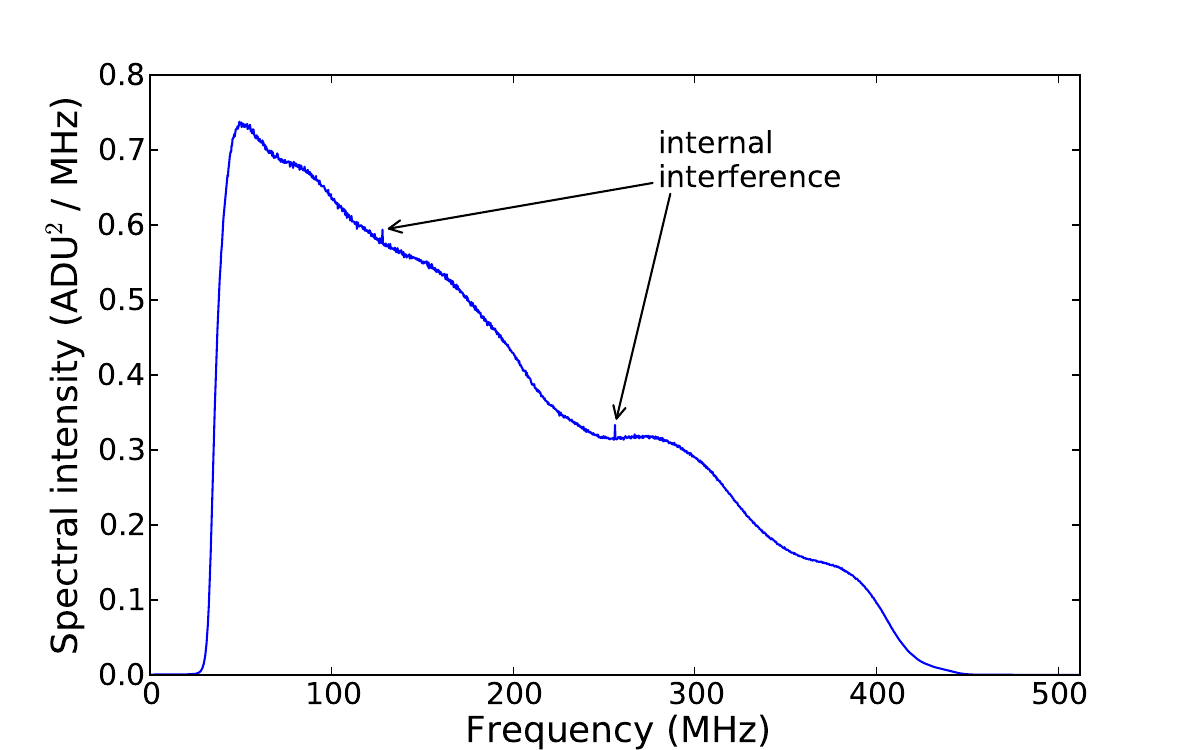}
 \caption{Power spectrum of the recorded dedispersion-filtered noise data.  The peaks at 128 and 256 MHz (marked) are internal interference from the digital signal-processing hardware, while the underlying spectrum results from the spectrum of the noise source and the bandpass of the dedispersion filter.}
 \label{fig:spect}
\end{figure}

A minor bug resulted in spurious recorded values at the start and end of each buffer, so these were excluded from the analysis; the triggering value, at the centre of the buffer, is unaffected by this.  Peaks in the spectrum at 0 and 512 MHz --- the bottom and top of the band --- were removed, the former being a direct-current offset and the latter probably resulting from internal interference from the digital hardware, at a harmonic of its clock speed of 256 MHz.  Lesser interference at harmonic and subharmonic frequencies was considered insignificant and not removed, but is just visible in the noise spectrum shown in figure~\ref{fig:spect}.

\subsection{Excursion rates}
\label{sec:exrate_bedlam}

\begin{figure}
 \centering
 \includegraphics[width=\linewidth]{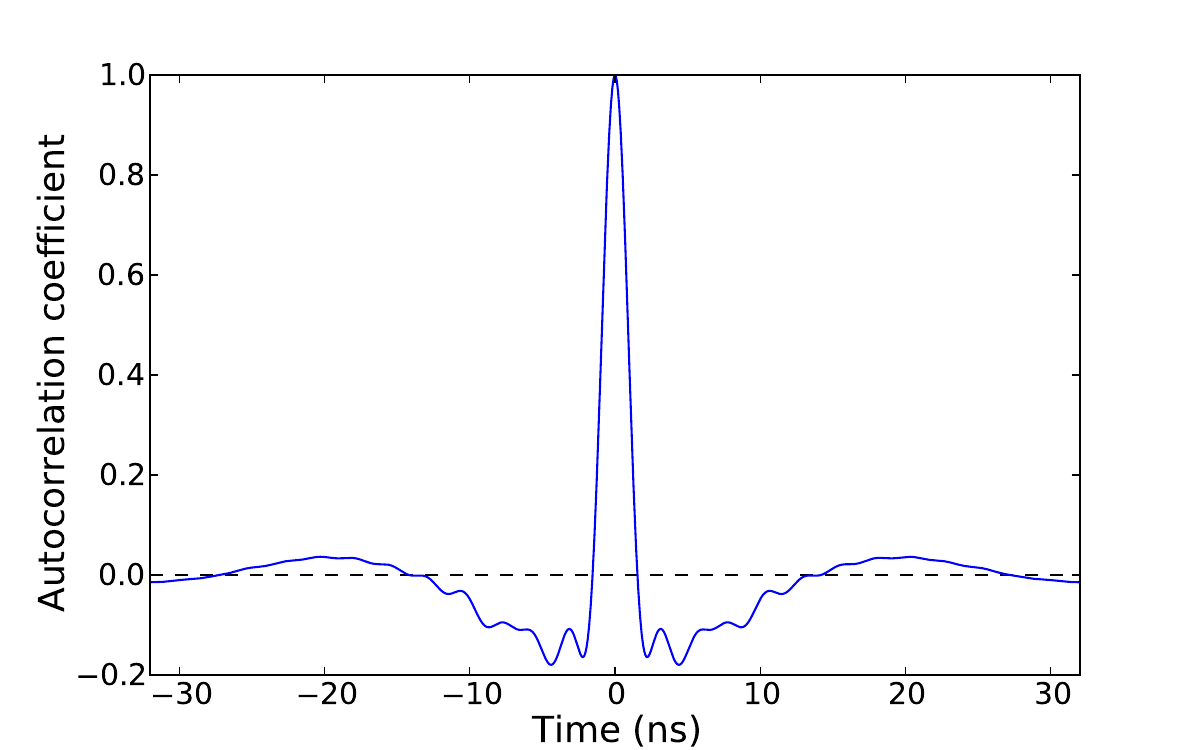}
 \caption{Autocorrelation function $\rho(t)$ for the recorded noise data around $t = 0$, derived from the power spectrum in figure~\ref{fig:spect}.}
 \label{fig:autocorr}
\end{figure}

Excursion rates for the signal and its envelope, with and without interpolation, were found as in section~\ref{sec:mc} for post-dedispersion buffers for a single channel of Bedlam.  Although the signal in this case does not undergo frequency mixing, unlike normal operation of the Parkes radio telescope (see figure~\ref{fig:sigpath}), random noise has no preferred phase and hence displays the same characteristics with or without this step.  For the same reason, the dedispersion filter has no effect on the characteristics of the signal other than to change its spectrum.  The spectrum after filtering is shown in figure~\ref{fig:spect} and the derived autocorrelation, for determining the expected excursion rates, is shown in figure~\ref{fig:autocorr}.

The significance of the peak in each buffer was calculated against a value of $\sigma$ equal to the combined RMS of all buffers within a moving 10-minute window, to reduce the uncertainty associated with the limited number of samples in a single buffer.  The central portion of each buffer, containing the triggering value, was excluded from the RMS calculation to avoid bias.  The RMS variation over a 10-minute period was $\sim 0.5\%$, resulting in a corresponding uncertainty in the peak height.

Quantisation error in the ADCs of $\pm 0.5$ ADU is equivalent to 0.5\% at 8$\sigma$.  Summed in quadrature with the uncertainty from RMS variation, above, this gives a 0.7\% uncertainty in the height of a single peak.  This has a negligible direct effect on the measured excursion rates for low threshold values, as it is averaged over a large number of excursions; it applies in full only to the single highest peak.  However, it indirectly causes a systematic effect: due to random increases or decreases in the peak height, the rate of excursions beyond a particular threshold will be a weighted average of the excursion rates that would occur at lower or higher thresholds in the absence of this variation.  This effect is equivalent to convolving the excursion rate as a function of threshold with the random error distribution.  Making a Gaussian approximation to the latter, we find that 0.7\% random error leads to a $\sim 0.2\%$ increase in the threshold at which a given excursion rate occurs.

\begin{figure}
 \centering
 \includegraphics[width=\linewidth]{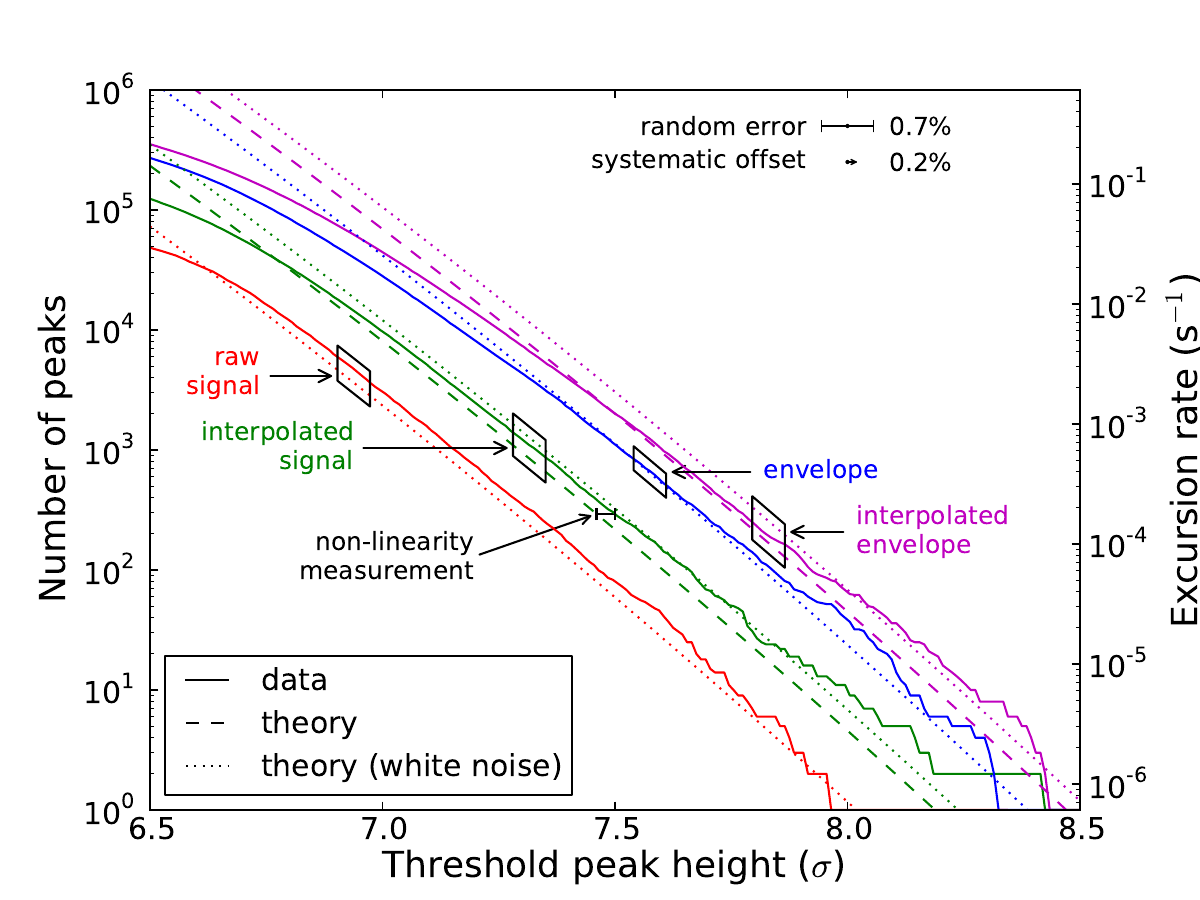}
 \caption{Measured excursion rates (solid) of the peak height in each buffer for experimental data with the Parkes Bedlam backend, for the signal and its envelope, with and without interpolation, compared with analytic predictions for the interpolated case (dashed; based on spectrum in figure~\ref{fig:spect}).  At low thresholds, there is a deficit in the excursion rates due to incomplete triggering.  At high thresholds, there is divergence due to random error and small-number statistics.  The systematic offset associated with the random error (see text) has already been applied to the predicted curves; we ascribe the remaining consistent excess at intermediate thresholds to non-linearity, which is measured as shown and discussed in section~\ref{sec:non-lin}.  Analytic predictions for the case of uncorrelated noise (dotted) are shown for comparison.}
 \label{fig:hist_bedlam}
\end{figure}

Figure~\ref{fig:hist_bedlam} shows the measured excursion rates compared to the predictions from equations~\ref{eqn:exrate_interpsig} and~\ref{eqn:exrate_interpenv}, modified by the above effect.  As in section~\ref{sec:mc_gauss} (figure~\ref{fig:hist_mc_gauss}), predictions for the case of uncorrelated white noise are also shown.  The results are consistent, in the sense that the difference between the excursion rates for the signal and for its envelope is clearly visible; but, unlike the Monte Carlo data in figure~\ref{fig:hist_mc_gauss}, these experimental data are not in sufficiently close agreement with the predicted excursion rates to discriminate against the alternative predictions for the case of white noise.

The deficit in the excursion rate at low thresholds is not caused by multiple excursions in a single array, as in the Monte Carlo data, but by the Bedlam backend failing to trigger on some lower-amplitude peaks.  As the triggering algorithm acts only on the signal (rather than its envelope), and performs only partial interpolation, a peak which is of low amplitude in the raw data but high amplitude after interpolation or in the envelope will be missed; consequently, the deficit extends to higher thresholds for interpolated and envelope data.  There is divergence at high thresholds from small-number statistics, as for the Monte Carlo data, but with an additional contribution from the random error described above.

The remaining discrepancy in figure~\ref{fig:hist_bedlam} is the small systematic excess at intermediate thresholds, most clearly visible for the raw and interpolated data.  Measuring this excess relative to the theoretical excursion rate for interpolated data at 7.5$\sigma$, as a compromise between uncertainty from incomplete triggering (at lower thresholds) and from small-number statistics (at higher thresholds), this is equivalent to a 0.4\% offset in the threshold.  The number of excursions beyond this threshold is 278, which is sufficient to make the averaged random error in the peak height negligible, but there is a remaining effect from small-number statistics: the Poisson error of $\pm 1 / \sqrt{278}$ in the number of excursions results in a $\pm 0.1\%$ uncertainty in the measurement of this offset.  We discuss the cause of this offset in the next section.

\subsection{Non-linearity}
\label{sec:non-lin}

In an ideal ADC, there is a direct linear relationship between the thresholds in the analog input voltage required to produce particular digital output codes, and the values of those output codes (in ADU).  In practice, there are minor offsets between these thresholds and their ideal values.   The maximum such offset for a given ADC is its integral non-linearity (INL); for the hardware used in this experiment, this is advertised as 0.4 ADU~\citep{e2v2010}.  This non-linearity affects measured peak heights, and hence excursion rates, through two different mechanisms.

The first effect is a direct error in the measurement of the peak height, with a magnitude less than or equal to the INL.  For this experiment, an INL of 0.4 ADU leads to a possible 0.4\% error for peaks at 7.5$\sigma$.  The direction and magnitude of the error may vary across the voltage range of the ADC, controlled by the difference in the non-linearity between adjacent voltage thresholds (the differential non-linearity, or DNL).  In general, we expect the error for peaks in the positive direction (around $+100$ ADU) to be uncorrelated with the error for peaks in the negative direction (around $-100$ ADU).

The second effect arises from mismeasurement of the RMS, which leads to an error in the normalised peak height (measured in $\sigma$).  This effect is dominated by the non-linearity in voltage thresholds within 1--2$\sigma$ (12--24 ADU) of zero voltage, within which most sample values lie.  The magnitude of this effect is difficult to predict, as it depends strongly on the variation of the non-linearity across this range, but we can place an approximate upper bound of the INL (0.4 ADU) divided by the RMS, or $\sim 3\%$, corresponding to minimal variation.  Because the same RMS value is used to normalise both positive and negative peak heights, we expect this error to apply to measured excursion rates in both directions.

The offset of $0.4 \pm 0.1\%$ in the threshold peak height for positive excursions in the first input channel of Bedlam, found in section~\ref{sec:exrate_bedlam}, is within the possible range of the above effects.  Repeating the analysis for negative excursions, we find an offset of $0.9 \pm 0.1\%$; the difference between these is explained entirely by the $\pm 0.4\%$ error arising directly from the peak height measurement.  Note that these offsets are both for the interpolated signal, rather than the envelope: as the envelope is defined to be positive, it neglects the distinction between positive and negative excursions.

\begin{table}
 \centering
 \begin{tabular}{ccc}
 \toprule
 \multirow{2}{*}{channel} & \multicolumn{2}{c}{non-linearity (\%)} \\
 \cmidrule{2-3}
  & positive & negative \\
 \midrule
 1A & $+0.4$ & $+0.9$ \\
 1B & $+0.3$ & $+1.0$ \\
 2A & $+1.7$ & $+2.5$ \\
 2B & $+1.1$ & $+1.2$ \\
 3A & $+0.5$ & $+0.4$ \\
 3B & $+0.8$ & $+1.0$ \\
 4A & $+0.8$ & $+1.6$ \\
 4B & $-0.4$ & $-0.3$ \\
 \bottomrule
\end{tabular}

 \caption{Offsets in threshold peak height, ascribed to non-linearity, required to produce the observed excess for rates of positive and negative excursions for interpolated data in all channels.  (Figure~\ref{fig:hist_bedlam} shows the measurement for positive excursions in channel 1A.)  The uncertainty in all measurements is $\pm 0.1\%$.  The channel labelling scheme reflects the normal usage of the Bedlam backend, with each channel connected to either the A or B polarisation of one of four beams of the Parkes 21 cm multibeam receiver.}
 \label{tab:nonlin}
\end{table}

Similar results were obtained for all input channels of Bedlam, each of which passes through a separate ADC.  The results are shown in table~\ref{tab:nonlin}, and are all within the $\pm 3\%$ range expected from RMS measurement error.  There is a clear correlation between the positive and negative excursion rates for each channel; the two highest and two lowest excesses are associated with channels 2A and 4B respectively.  This is consistent with the greater part of the common excess originating from error in the RMS measurement, while the difference between the excesses for positive and negative excursions is caused directly by errors in measurement of the peak height.  The greatest such difference is 0.8\% for channel 2A, which is just compatible with independent errors of $\pm 0.4\%$ in measurement of the peak height, combined with $\pm 0.1\%$ from measurement of the excess in the excursion rate.

\section{Conclusion}

Experiments which search for radio transients with a time-scale comparable to the inverse bandwidth, such as attempts to detect neutrino-induced particle cascades in the Moon, are in a different regime to transient searches on longer time-scales and subject to different effects.  In particular, the peak amplitude of a pulse is not generally recovered in the sampled data, and the process of mixing the signal to shift it in frequency can reduce the peak amplitude further.  In such an experiment it is necessary to compensate for these effects, and to understand the statistics of the resulting signal, to evaluate the significance of a possible detection.

The process of sampling can be reversed by interpolating between the sampled values.  As frequency mixing causes a loss of phase information, it is not possible to unambiguously reconstruct the original signal; but the signal envelope can be computed to recover the peak amplitude of a pulse, at the cost of increasing the amplitude of the noise.  We have derived the expected excursion rate from Gaussian noise after these steps, and confirmed it through simulation and experiment.  We have also described and tested Bedlam, a digital backend for the Parkes radio telescope which is designed for such an experiment.

\begin{acknowledgements}
The Parkes radio telescope is part of the Australia Telescope which is funded by the Commonwealth of Australia for operation as a National Facility managed by CSIRO. This research was supported by the Australian Research Council's Discovery Project funding scheme (project number DP0881006).
\end{acknowledgements}

\bibliographystyle{splncsnat}
\bibliography{allbib}

\end{document}